\documentclass[galaxies,article,submit,moreauthors,pdftex,10pt,a4paper]{mdpi} 
\preto{\abstractkeywords}{\nolinenumbers}

\firstpage{1} 
\articlenumber{x}
\doinum{10.3390/------}
\pubvolume{xx}
\pubyear{2017}
\copyrightyear{2017}
\externaleditor{Academic Editor: name}
\history{Received: date; Accepted: date; Published: date}

\Title {\uppercase{Detection of Periodic radio signal from the blazar J1043+2408}}

\Author{Gopal Bhatta $^{1,\dagger,\ddagger}$\orcidA{0000-0002-0705-6619}*}

\AuthorNames{Firstname Lastname, Firstname Lastname and Firstname Lastname}

\address{%
$^{1}$ \quad Astronomical Observatory, Jagiellonian University, ul. Orla 171, 30-244 Krak\'ow, Poland; gopalbhatta716@gmail.com\\
}
\corres{Correspondence: gopalbhatta716@gmail.com; Tel.: +x-xxx-xxx-xxxx}

\abstract{ Search for periodic signals from blazars has become widely discussed topic in recent years. In the scenario that such periodic changes originate from the innermost regions of blazars, the signals bear imprints of the processes occurring near the central engine, which is mostly inaccessible to our direct view. Such signals provide insights into various aspect of blazar studies including disk-jet connection, magnetic field configuration and, more importantly, strong gravity near the supermassive black holes and release of gravitational waves from the binary supermassive black hole systems. In this work, we report detection of a periodic signal in the radio light curve of the blazar  J1043+2408  spanning $\sim$10.5 years. We performed multiple methods of time series analysis, namely, epoch folding, Lomb-Scargle periodogram, and discrete auto-correlation function. All three methods consistently reveal a repeating signal with  a periodicity of $\sim560$ days.  To robustly account for the red-noise processes usually dominant in the blazar variability and other possible artifacts, a large number of Monte Carlo simulations were performed. This allowed us to estimate a high significance (99.9\% local and 99.4\% global) against possible spurious detection. As possible explanations, we discuss a number of scenarios including binary supermassive black hole system, Lense-Thirring precession and jet precession.} 


\begin{document}
%%%%%%%%%%%%%%%%%%%%%%%%%%%%%%%%%%%%%%%%%%
%% Only for the journal Gels: Please place the Experimental Section after the Conclusions

%%%%%%%%%%%%%%%%%%%%%%%%%%%%%%%%%%%%%%%%%%

\section{Introduction}
Blazars, a class of radio-loud active galactic nuclei (AGN), are the most energetic sources in the universe. The sources have relativistic jets  beamed upon us that shine brightly in non-thermal emission covering a wide electromagnetic spectrum - from radio to most energetic $\gamma$-rays.  As jets accelerate matter with high speeds down the jet orientated close to the observer's line of sight, relativistic effects become dominant resulting in the Doppler boosted emission that is highly variable over the entire electromagnetic spectrum \citep{Meier12}. The broadband spectral energy distribution (SED) of blazars can often be identified with a double-peaked feature in the frequency-flux plane. The lower peak, usually found between the radio and X-ray, arises as a result of the synchrotron emission by the energetic particles accelerating in the jet magnetic field;  whereas the origin of the high frequency component, extending from X-rays to TeV energies, is still a debated topic. In leptonic models, e.g., \cite{Maraschi1992,Bloom1996} it is resulted due to the inverse-Compton scattering of the soft seed photons by the energetic particles. In the synchrotron self-Compton scenario (SSC), the same population of the high energy electrons responsible for the synchrotron emission up-scatter the photons to higher energy. However, in the external Compton scenario (EC), the seed photons might originate at the various components of an AGN, e.g., accretion disk \cite{Dermer1993}, broad-line region  \citep[][]{Sikora1994}, and dusty torus \cite{Blazejowski2000}.  Hadronic models, on the other hand, ascribe the high energy emission from blazars to the interaction of relativistic protons in the presence of radiation fields \citep[e.g.,][]{Mannheim1992,Aharonian2000,Mucke2003}

Blazars consist of two types of sources: flat-spectrum radio quasars (FSRQ) and  BL Lacertae (BL Lac) objects. FSQRs are the more powerful sources which show emission lines  in the optical continuum and have the synchrotron peak in the lower part of the spectrum; whereas BL Lac objects are less powerful ones which show weak or no emission lines and have synchrotron peak in the higher part of the spectrum. BL Lacs represent an extreme class of sources with maximum synchrotron emission output reaching up to soft X-rays (10$^{16}$ Hz) and inverse-Compton emission output ranging from hard X-rays to TeV emission. The sources do not possess strong circum-nuclear photon fields and are believed to accrete at relatively low rates \cite[see][and references therein]{Bhatta2018b}.

Blazars display variability in a wide range of temporal frequencies - equivalently, on diverse timescales ranging from a few minutes to decades.  The statistical nature of such variability can fairly be represented by a featureless power-law power spectral density (PSD) \cite[see in this context][and references therein]{Bhatta2016,Bhatta2018c}. However, signatures of quasi-periodic oscillations (QPO) in the multi-frequency blazar  light curves, including radio, optical, X-ray and $\gamma$-ray, have been found. The timescales of the reported periodicity range from a few hours to a few years \cite[see][for QPOs in blazars]{Gupta2018,Bhatta2017,Zola2016,Bhatta2016b}. 

 In particular,  at radio frequencies QPO on diverse timescales have been recorded  in a number of blazars: In BL Lac source PKS 0219-164,  a strong signal of QPO with 270 d period, 
along with possible low-frequency harmonics, was detected  \cite{Bhatta2017}.  Similarly, in BL Lac source AO\,0235, QPOs with periods ranging from about one year up to several years  have been reported  (\cite{Liu2006,Raiteri2001}. Also, FSRQ J1359+ 4011 was reported to exhibit persistent $\sim 150$\,day periodic modulation in the 15\,GHz observations \cite{King2013}.  In addition, several other blazar sources were found to show QPOs in the radio frequencies, e.g.,  FSRQ PKS\,1510-089 \citep{Xie2008}, blazar NRAO\,530 \citep{An2013}, and FSRQ PKS\,1156+295 \cite{ Wang14,Hovatta2008, Hovatta2007}.

 \begin{figure*}[t!]
\centering
\includegraphics[width=0.98\textwidth]{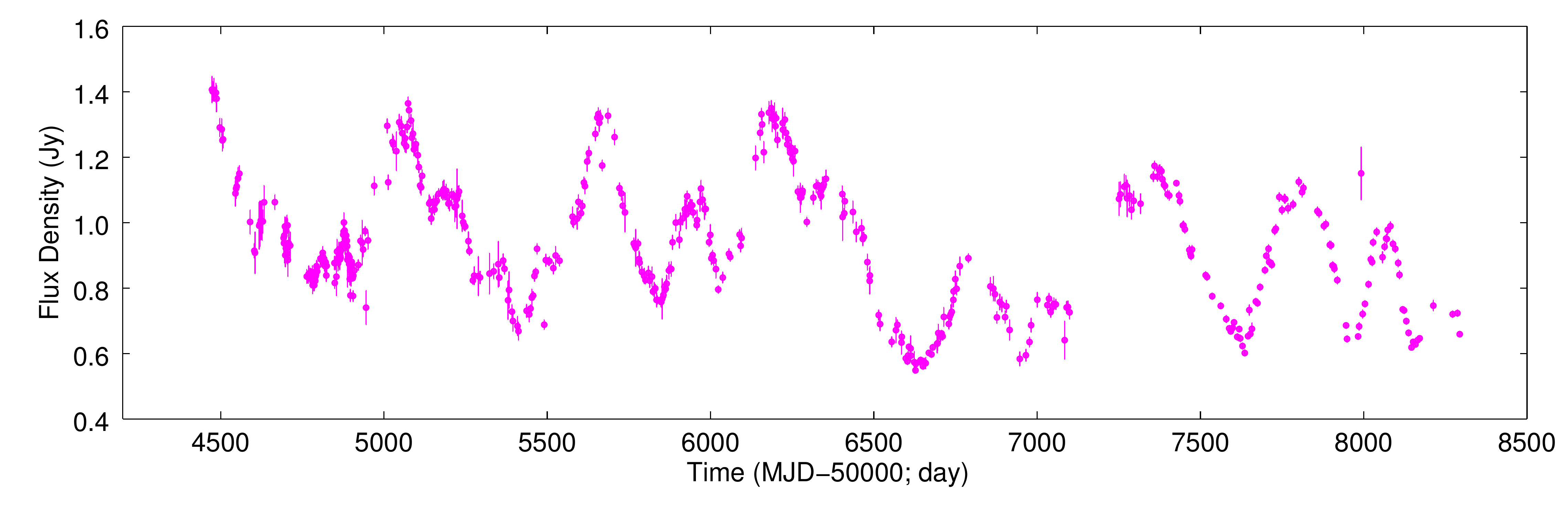}
\caption{$\sim10.5$ years long 15 GHz  observations of the blazar J1043+2408 from OVRO}
\label{LCs}
\end{figure*}

BL Lac J1043+2408 (RA=10h\,43m\,09.0s, Dec= +24d\,08m\,35s, and z = 0.563446; \cite{Hewett2010}) has been detected by most of the currently available instruments operating within a wide range of electromagnetic frequencies. The source is cataloged by Fermi/LAT as 3FGL J1043.1+2407 \cite{Acero2015}, and  in the soft X-ray (0.1-2.4 keV) it was frequently observed by ROSAT \cite{Massaro2009}. During The Micro-Arcsecond Scintillation-Induced Variability Survey III, its  optical (R-band)  brightness was recorded to be 16.84 magnitudes \cite{Pursimo2013}. The blazar is being regularly monitored  in the 15-GHz radio band by the 40-m telescope of the Owens Valley Radio Observatory (OVRO) since 2008.

 In this paper, we analyze the long term ($\sim 10.5$ years) radio observations of the blazar J1043+2408 and report detection of 563 d periodicity in the light curve. In section 2 we discuss data acquisition.  We present time series analysis of the light curve using the epoch-folding, Lomb-Scargle periodogram (LSP), and discrete auto-correlation function methods; besides, we also elaborate on the Monte Carlo simulation technique which is used to compute the statistical significance of the detected periodicity.  In section 4, we discuss various possible scenarios that can lead to the observed periodic signal; and finally we summarize our conclusion in section 5.
 
 %%%%%%%%%%%%%%%%%%%%%%%%%%%%%%%%%%%%%%%%%%
\section{ Data Acquisition}
\label{sec:obs}

The 15 GHz radio observations of the source J1043+2408 were obtained from Owens Valley Radio Observatory (OVRO; \cite[][]{Richards2011}). We analyzed the observations, with an average sampling of a week, from epoch 2008-01-08 to 2018-07-08 (equivalently, from MJD 54473 to 58307), spanning 3820 days ($\sim$ 10.5 years) .

\section{Analysis and Results}
 The 15 GHz band  light curve  of the blazar J1043+2408  from the last decade is presented in Figure \ref{LCs}.  In the figure, it can be clearly seen that not only the source displays long-term variability but it also shows a periodic flux modulation such that in each cycle the flux nearly doubles between minimum and maximum. To quantify the observed  long-term variability, we estimated a fractional variability \cite[][]{Vaughan2003,Bhatta2018a} of $20.29\pm0.63$ \% , indicating a moderate variability over the period.
%%%%%%%%%%%%%%%%%%%%%%%%%%%%%%%%%%%%%%%%%%

\subsection{Periodicity Search: }
We carried out periodicity search analysis using three well known methods of time-series analysis: epoch folding, Lomb-Scargle periodogram and discrete auto-correlation function. The methods, analyses and the results are discussed in detail below.

 \begin{figure}[t!]
\centering
\begin{tabular}{c@{}c}
%\begin{tabular}{@{}p{6.5cm}@{}@{}p{6cm}@{}}
\hspace{-0.48cm}
\resizebox{0.52\textwidth}{!}{\includegraphics{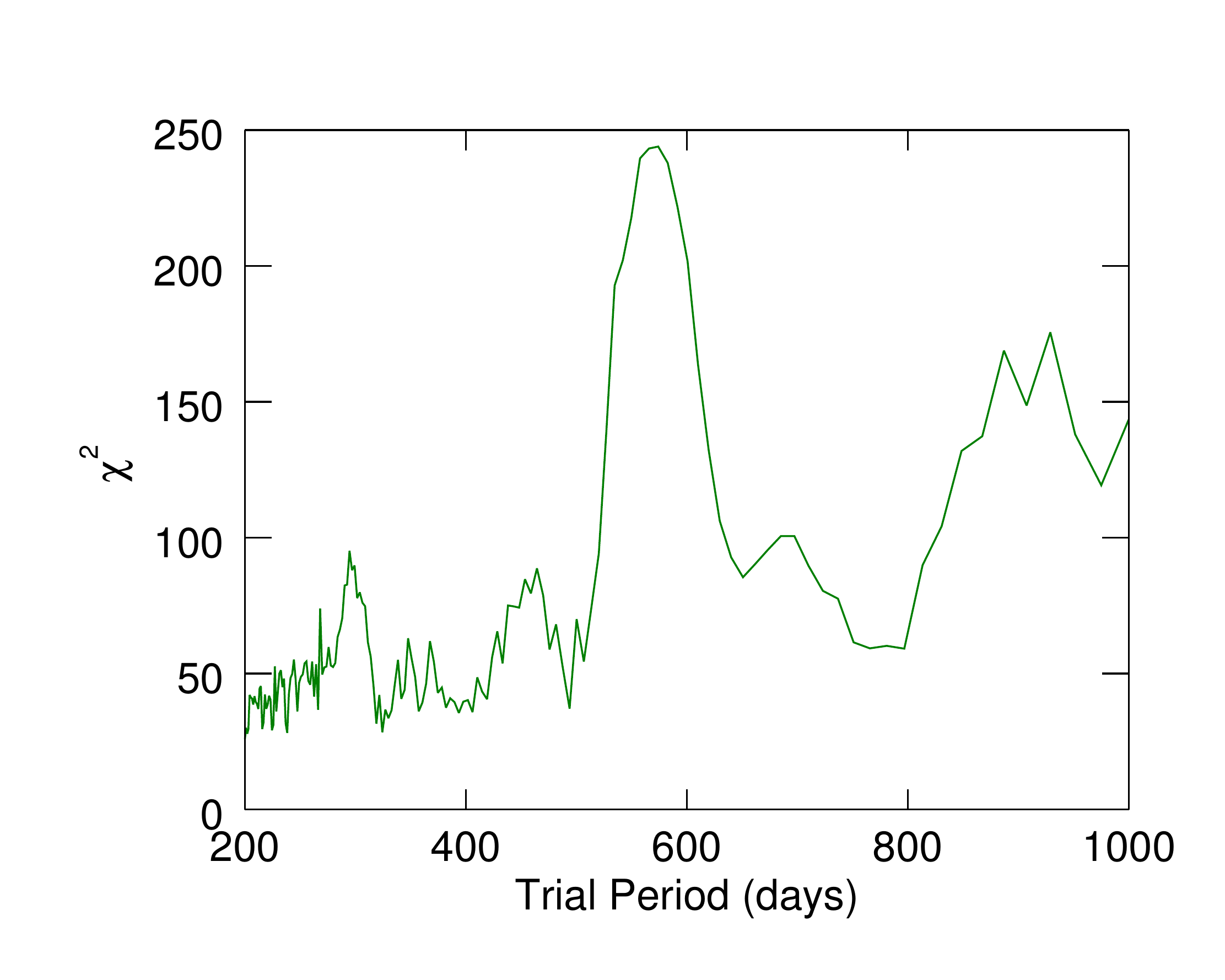}} &
\hspace{-0.48cm}
\resizebox{0.52\textwidth}{!}{\includegraphics{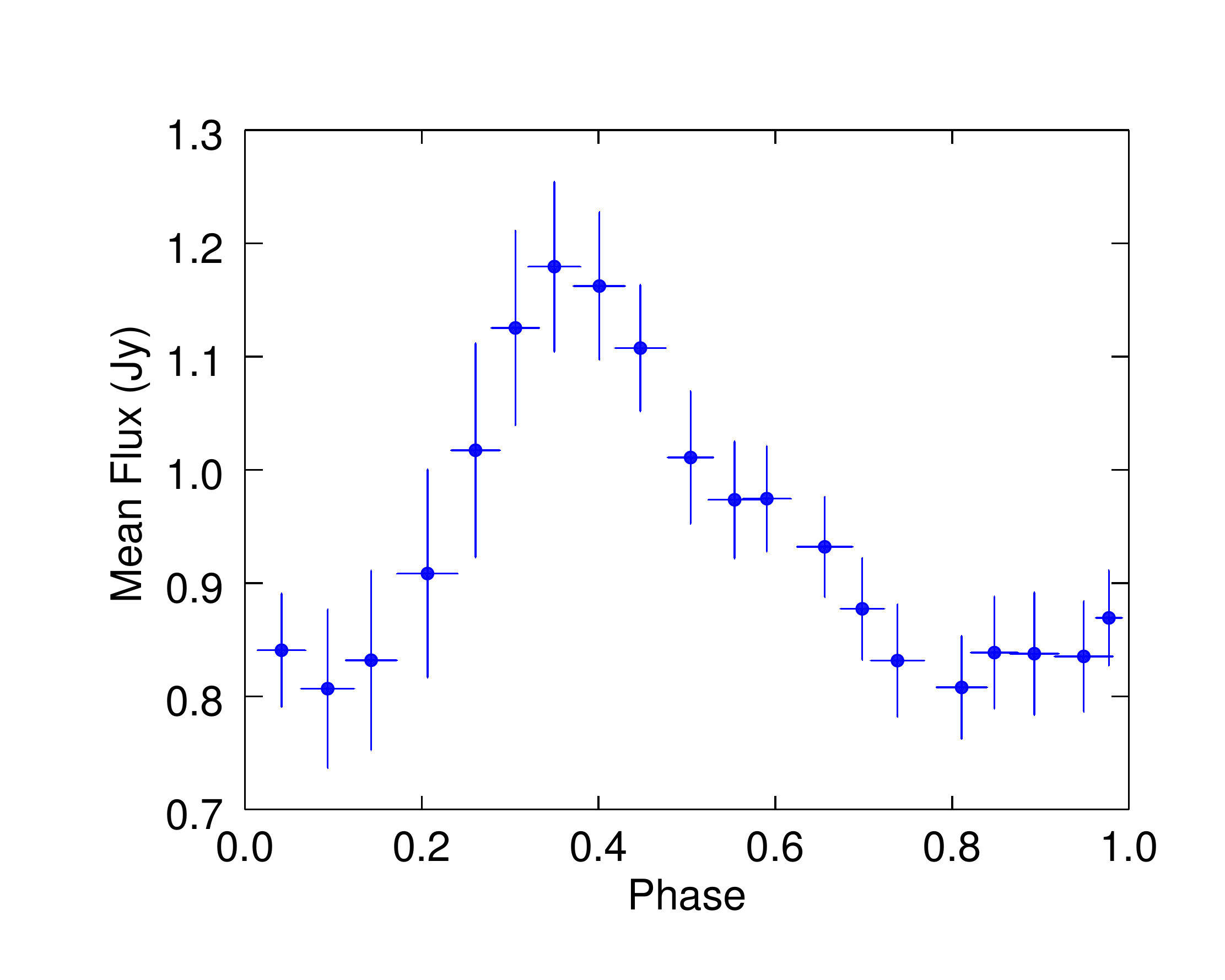}} \\
\end{tabular}
\caption{Epoch folding of the radio light curve of the blazar J1043+2408. Left panel: $\chi^2$ values for the trial periods between 200 and 1000 days. Right panel: pulse profile of the periodic component at the timescale of 563 days representing maximum $\chi^2$ value.}
\label{fold}
\end{figure}

\subsubsection{Epoch folding} 
 Epoch folding,  a widely discussed method of time series analysis, was first worked out  by Leahy et al. (1983) \cite{Leahy1983} and later improved by a number of authors \cite[e.g.][]{Davies1990,Davies1991}.  Unlike traditional discrete Fourier periodogram, which expects periodic components to be of the sinusoidal shape, the method is less sensitive to the modulating shape of the periodic components. Also, this method is largely unaffected by the irregularity in the sampling of the observations and therefore well suited for the periodic search in the data with gaps. 
 In this method,  a time series with N data points and varying about a mean value of $\bar{x}$ is folded on several trial periods and phase bins. Then, a quantity $\chi ^{2}$ expressed as
\begin{equation}
\chi ^{2}=\sum_{i=1}^{M}\frac{\left (x_{i}-\bar{x}\right )^2}{\sigma_{i}^{2}}, 
\label{chisq}
\end{equation}
\noindent is estimated, where $x_{i}$ and $\sigma_{i}$ represent the mean and standard deviation, respectively, of each of M phase bins. For the observations distributed as Gaussian noise, we generally find $\chi ^{2}$ $\sim M$. However in case of the observations containing periodic signals, $\chi ^{2}$ takes a value which is significantly different (or the maximum) from the average value \cite[for details refer to][]{Larsson1996}. The method has been frequently tested in the context of the periodicity analysis of blazar light curves e.g., \cite{Zhang2014}.

We computed $\chi ^{2}$ values of the source light curve for the trial periods ranging between 200 and 1000 days using a time step of 14 days. The  pulse profiles corresponding to these trial periods were generated and, subsequently, tested for $\chi^2$ constancy using  Equation \ref{chisq}. The left panel of Figure \ref{fold} presents the distribution of the $\chi^2$ values over the trial periods considered. The maximum $\chi^2$  deviation seen at $563\pm49$ days, represents the most probable period. The uncertainties here represent the half-width at maximum (HWHM) of the Gaussian fit of the peak centered around 563-day trial period. The right panel of Figure \ref{fold} shows the pulse profile corresponding to the period. The observed period is further tested by using the following methods.

  \begin{figure}[t!]
\centering
\begin{tabular}{c@{}c}
\hspace{-0.5cm}
\resizebox{0.52\textwidth}{!}{\includegraphics{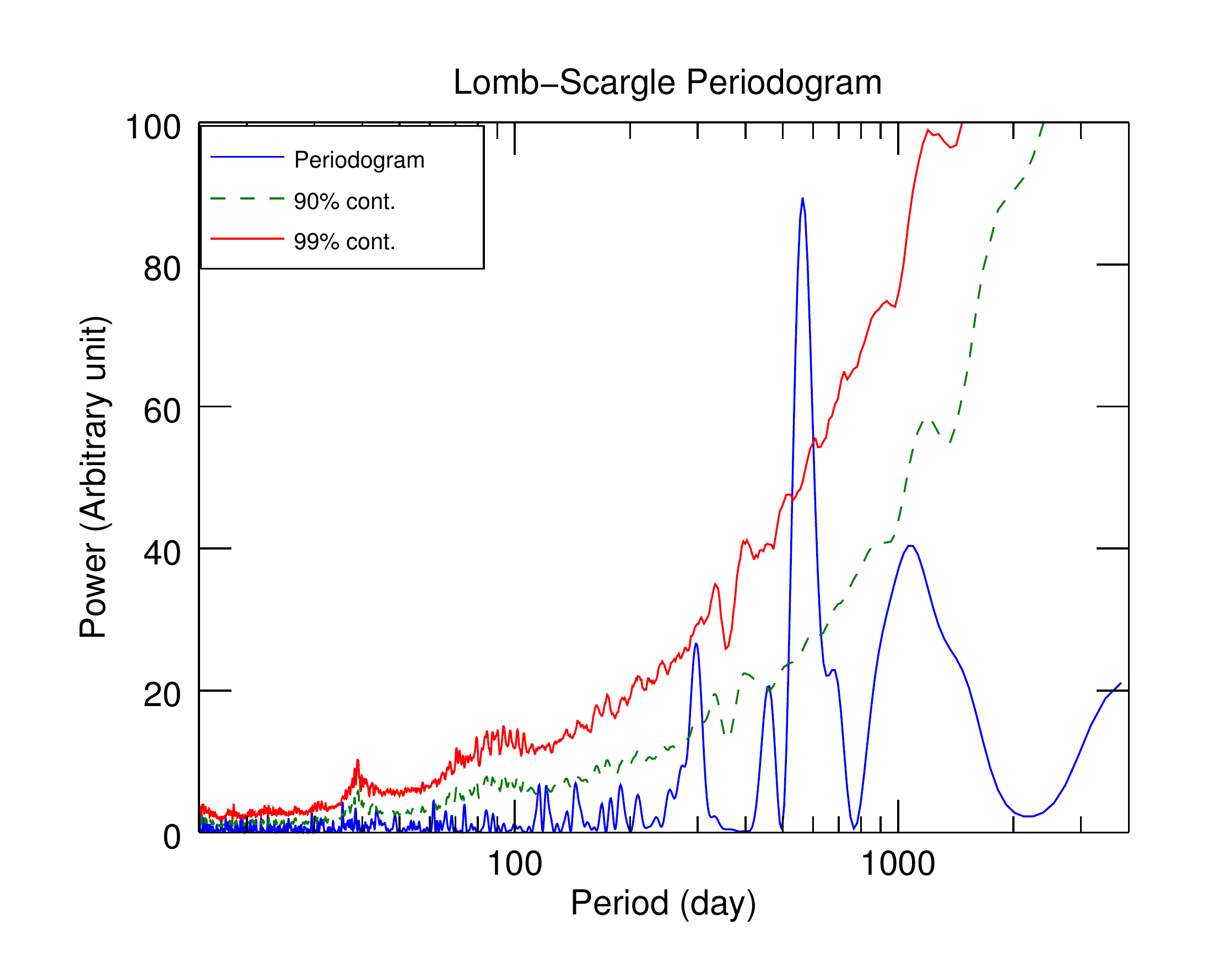}} &
\hspace{-0.68cm}
\resizebox{0.52\textwidth}{!}{\includegraphics{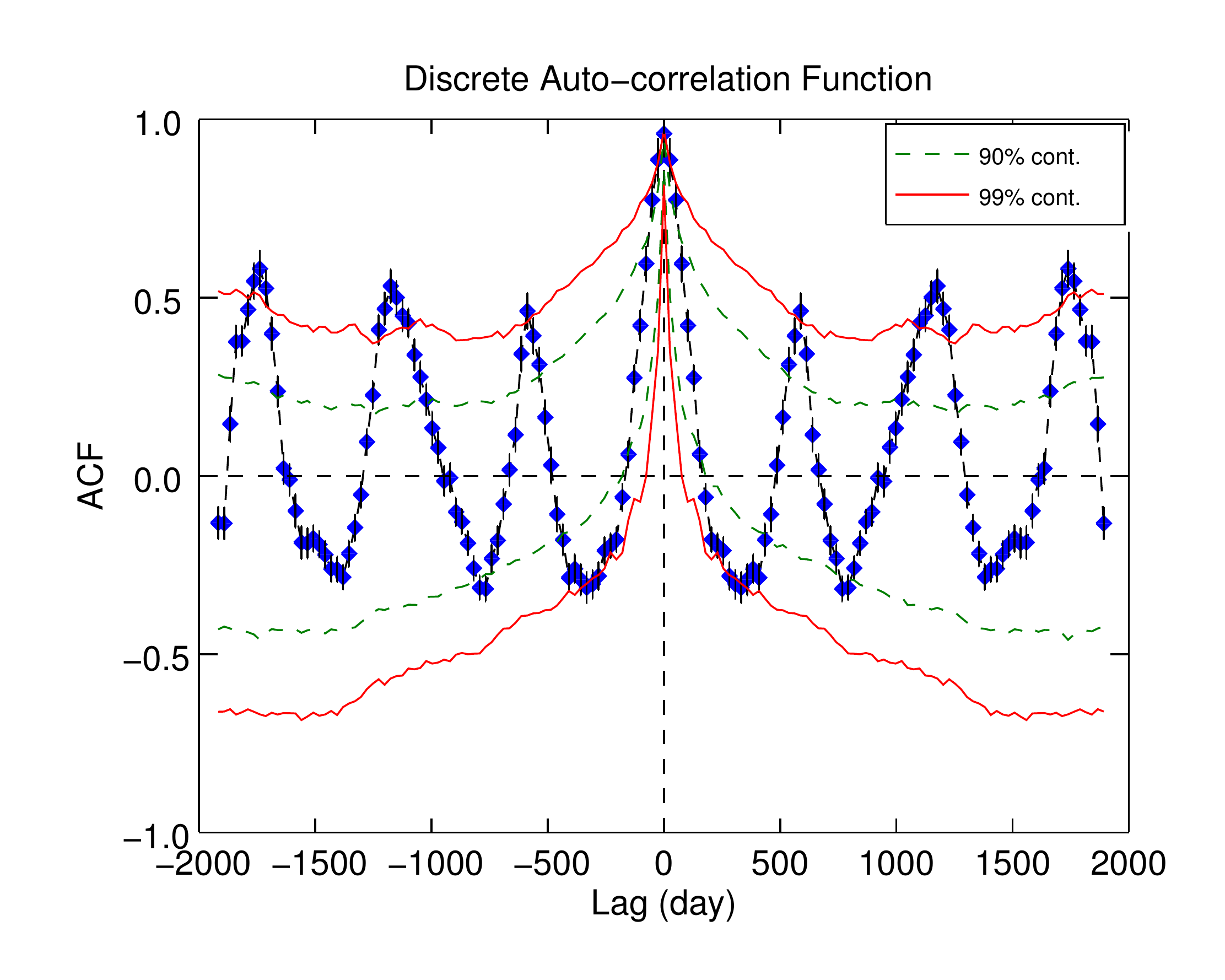}} \\
\end{tabular}
\caption{Lomb-Scargle periodogram (left panel) and discrete auto-correlation function (right panel) of the $\sim10.5$ years long 15 GHz observations of the blazar J1043+2408. The dashed green and red curves represent the 90 and 99\% significance levels, respectively, from MC simulations.}
\label{LSP}
\end{figure}
                          
\subsubsection{Lomb-Scargle Periodogram} 
The Lomb-Scargle Periodogram \cite[LSP;][]{Lomb1976,Scargle1982}  is one of the most popular methods of time series  analysis \cite[see][]{Bhatta2016,Bhatta2017,Bhatta2016b}. The method, as a form of least-square fitting of the sinusoidal waves to the data, is less sensitive to the gaps in the data when compared to the traditional discrete Fourier transform (DFT). The fitting process enhances the periodogram features (or peaks) that can possibly represent periodic signals in light curves.  The  LSP of the source light curve was computed for the minimum and maximum frequencies of, $f_{min} = 1/3820$ d, and  $ f_{max}$=1/14 d, respectively.
 It should be pointed out that the choice of the total number of periodogram frequencies, $n_{0}$, plays an important role in the evaluation of the periodogram; we would like to select its value keeping in mind that a small value lacks  precision on the location of the most prominent period; whereas a large value could be computationally inconvenient.
 In this work, the total number of periodogram frequencies were evaluated using  
\begin{equation}
N_{eval}=n_{0} T f_{max} \rm ,
\label{nf}
\end{equation}
\noindent  where $n_{0}=10$, and $T = 3820$ days represents the total length of the observations  \citep[see][]{VanderPlas2018}.  The LSP of the $\sim$10.5 years long OVRO  light curve of the  blazar J1043+2408 is presented in Figure~\ref{LSP}. It can be seen that in the periodogram a distinct peak stands out around the timescale of $563\pm42$ days; this suggests presence of a strong periodic signal at the timescale. The uncertainties in the period are estimated by taking the HWHM of the most prominent peak \cite[see][]{Bhatta2017,Bhatta2016b}.  Moreover, to assess the effect of $n_{0}$, we also computed LSP using $n_{0}=4$. This made the dominant period shift to $567$ days with larger uncertainties of $47$ days. We see that the results are consistent with the previous finding by the epoch folding method.

\subsubsection{Discrete auto cross-correlation Function}
To further confirm the presence of the above periodic timescale using different method, we performed the discrete correlation function analysis as described in \cite{EK88}. The method has been mostly applied to investigate  cross-correlation between two time series with uneven spacing  \citep[see][]{Bhatta2018b, Bhatta2018c,Bhatta2016,Bhatta2018a}. With only one light curve, the method becomes discrete auto-correlation function (ACF), which can be exploited to reveal the periodic signals in a light curve. The ACF, although related to  PSD in frequency space, is computed in time domain and therefore it is free of the sampling effects such as windowing and aliasing. The method has been frequently employed in the periodicity search in  blazar light curves \cite[e.g.][]{Bhatta2018b,Villata2004}. The discrete ACF of the radio light curve of the blazar is presented in the right panel of  Figure \ref{LSP} which clearly reveals presence of the periodic behavior represented by the ACF peaks recurring after an interval of $563\pm46$ days. Once again, HWHM for the first peak is used as a measure of uncertainty. It should be noted that all three estimates of uncertainty measures, as provided by the HWHM of the Gaussian fit of the most prominent features, are consistent within 2\% of the significant period. 

\subsubsection{Significance Estimation and Monte Carlo Simulation}
After we have employed three different methods to detect periodicity in the light curve, it should be pointed out that, in general, the statistical properties of blazar light curves can be characterized as red-noise processes that can potentially mimic a transient periodic behavior, especially in the low-frequency domain \citep[see][for the discussion]{Press1978,Bhatta2017,Bhatta2018c}. In addition, spurious peaks might arise owing to other sampling effects  including discrete sampling, finite observation period and uneven sampling of the light curve.  Therefore, it is important to make consideration of these effects in the significance estimation against spurious detection. 
 To address the issue, we followed the power response method  \citep[PSRESP;][]{Uttley2002}, a  method extensively used in the characterization of the AGN power spectrum density \citep[see][and references therein]{Bhatta2016,Bhatta2016b,Bhatta2017,Bhatta2018b}. 
  First the source periodogram was modeled with a power-law PSD of the form $P(\nu) \propto \nu ^{-\beta}+C$; where ${\nu}$, ${\beta}$, and $C$ represent temporal frequency, spectral slope and Poisson noise level, respectively.  To maximize the probability that the PSD model best represents the observed periodogram, $\beta=1.3\pm0.1$ and $C=2.7$ were used.  A large number of (typically 10000)  light curves were then simulated following the Monte Carlo (MC) method described in \cite[][]{Timmer1995}. The method fully randomizes both the amplitude and the phase. However, the method could have possible caveats due to the fact that it produces Gaussian distributed light curves, which may not represent the probability density function of the observations \cite[see][]{Emmanoulopoulos2013}. Nevertheless, as the periodic features in our all methods clearly stand out, we expect that any significant deviations in the significance estimation will fall within the range of uncertainties. 
  
  The simulated light curves were re-sampled to match the sampling of the source light curve.   
Now to estimate the  significance of the periodic feature seen in the LSP, the LSP distribution of the simulated light curves was utilized to estimate a local and global significance of the periodic feature against spurious detections \citep[for further details see][]{Bhatta2016,Bhatta2016b,Bhatta2017,Bhatta2018b}. The local significance tells us how likely the observed feature at a particular frequency is significant; whereas the global significance, which considers LSP distribution at all the frequencies considered, accounts for the fact that we do not have an {\it a priori} knowledge of the location of the frequency at which the most significant feature might occur \cite[see][]{Bhatta2017}. The local significance of the observed periodic feature at the period of $\sim$563 d turned out to be $\sim$99.4\%, and similarly the global significance of the LSP peak at the period $\sim$563 day was evaluated to be 99.6\%.  The local 90\% and 99\% LSP significance contours are represented by the dashed green and the red curves, respectively, in the left panel Figure~\ref{LSP}. It can be seen that the red-noise has a predominance of spectral powers in the low-frequency regime, or equivalently longer timescales.

 In similar way, we used the distribution of the simulated ACFs to estimate the local 90\% and 99\% significance contours shown by the dashed green and the red curves, respectively, in the auto-correlation function shown in the right panel of Figure~\ref{LSP}. The observed high significances imply a low probability of spurious detections and therefore suggests that the signal, intrinsic to the source light curve, should have a physical origin.

%%%%%%%%%%%%%%%%%%%%%%%%%%%%%%%%%%%%%%%%%%
\section{Discussion}
Study of periodic oscillations in blazars could be a novel method to investigate the processes occurring at the innermost  regions of the active central engines. The studies could provide important insights into a number of blazar aspects including strong gravity environment around fast spinning supermassive black holes (SMBH), magnetic field configuration near the accretion disk, disk-jet connection and release of gravitational waves (GW) from the binary supermassive black hole systems. There could be a number of processes that can explain the observed periodic flux modulations.  Below we discuss some of them.

\begin{figure}[]
\begin{center}
\begin{tabular}{c@{}c}
\hspace{-0.50cm}
\resizebox{0.52\textwidth}{!}{\includegraphics{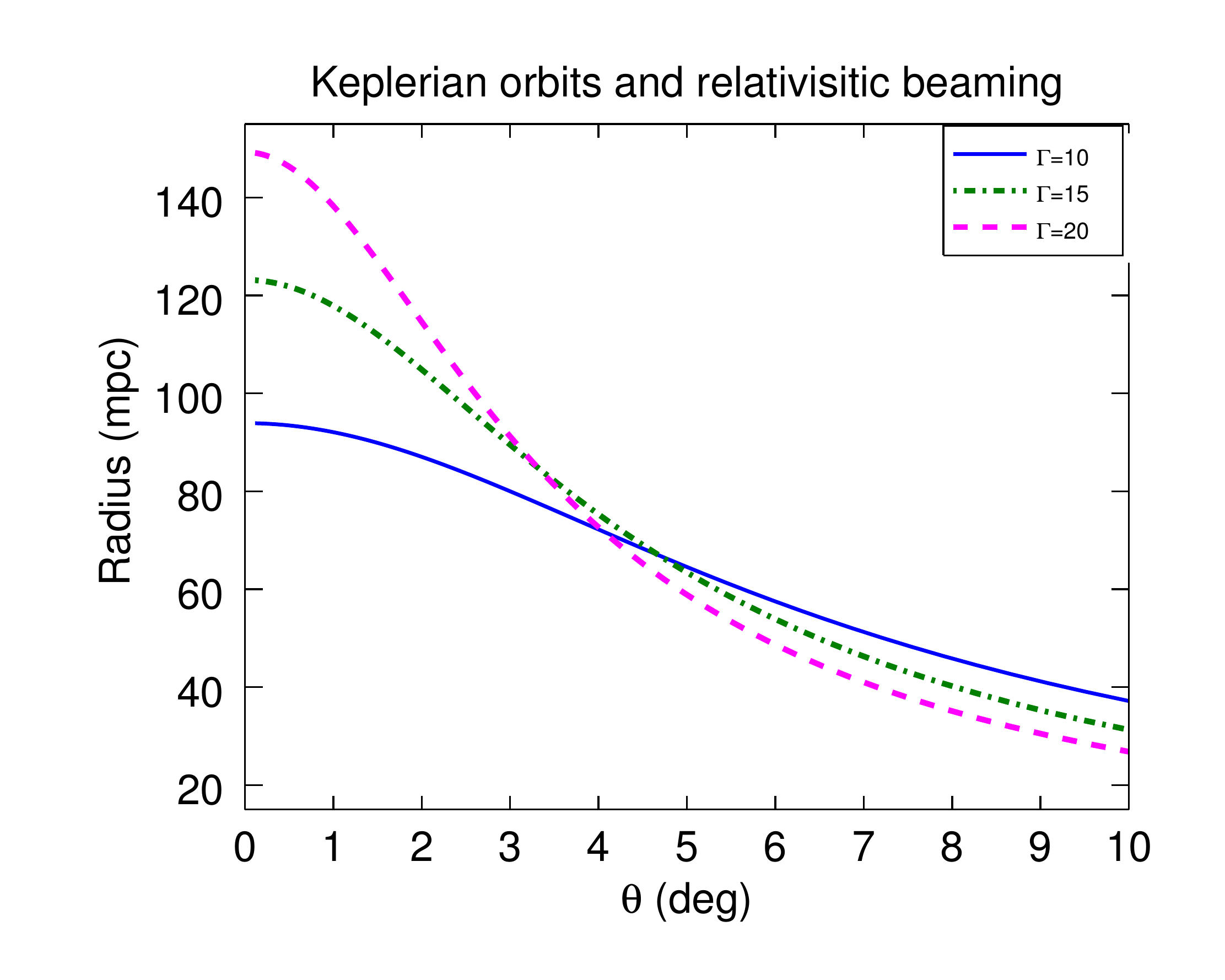}} & 
\hspace{-0.58cm}
\resizebox{0.52\textwidth}{!}{\includegraphics{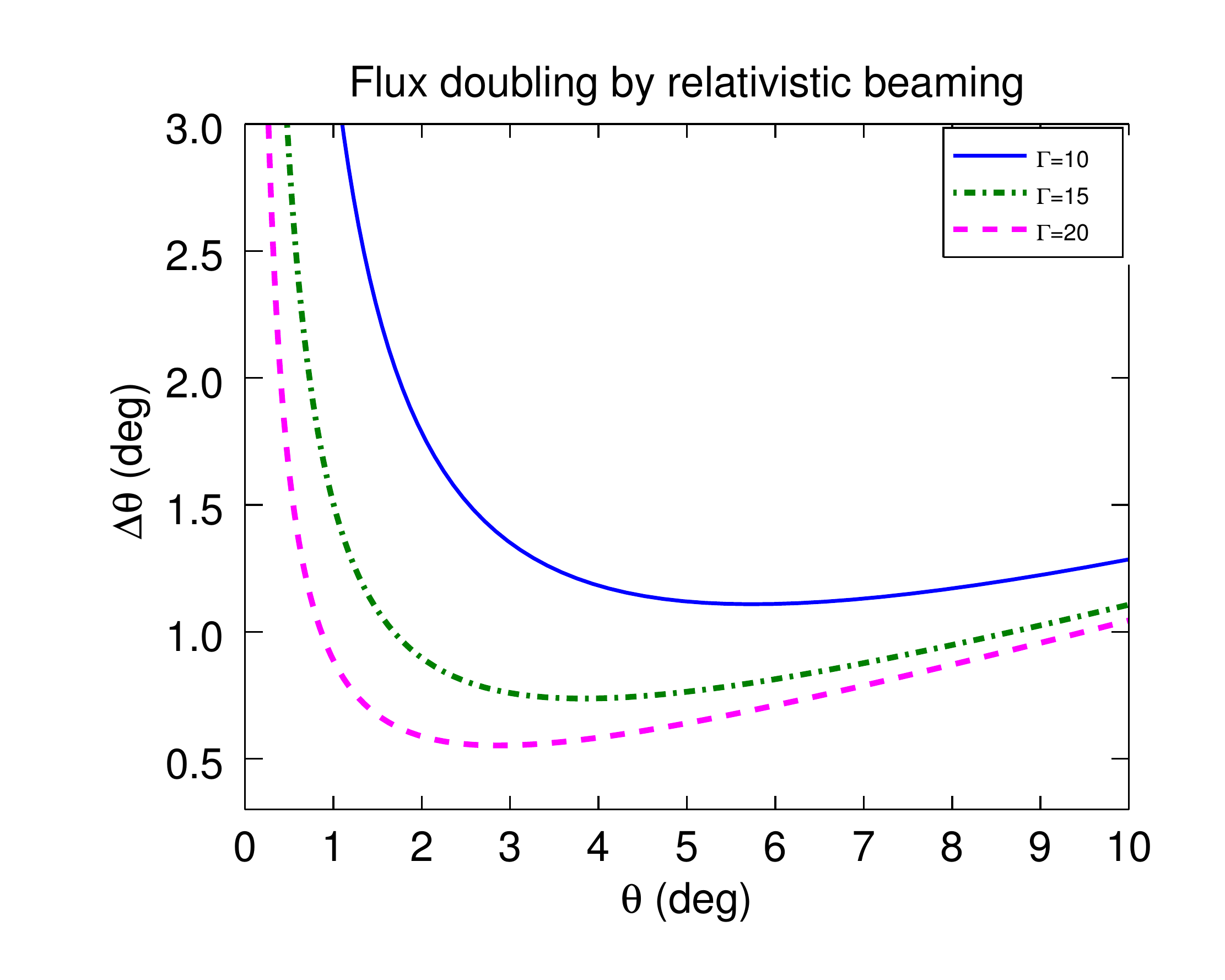}}\\
\end{tabular}
\caption{Left panel: Change in the radii (in milli-parsecs; mpc) of the Keplerian circular orbits around the black hole of mass of $10^9M_\odot$ corresponding to the 563 d periodic signal carried by the emission regions traveling with the bulk Lorentz factors, $\Gamma =10, 15$, and 20, along the blazar jet viewed close to the line of sight. Right panel: The change in the viewing angle as a function of viewing angle for $\Gamma =10, 15$, and 20 required to double the apparent flux, keeping intrinsic flux constant, via relativistic beaming. A radio spectral index $\alpha$=0.6, typical of blazars, is used here.  \label{Fig4}}
\end{center}
\end{figure}

First, for the observed period ($P_{obs}$) of 563 days, the corresponding period in the source rest frame ($P$) at $z=0.563$  is 360 days, as estimated using $P=P_{obs}/(1+z)$. If we take the value as the Keplerian period around the central black hole, $\tau_k$, we can  estimate the corresponding radius of the orbit as given by
\begin{equation}
\tau_k=0.36 \left ( \frac{M}{10^9M_\odot} \right )^{-1/2}\left ( \frac{a}{r_g} \right )^{3/2} \rm days,
\end{equation}
where $a$ is the length of the semi-major axis of the elliptic orbits. For a black hole of mass of $10^9 M_\odot$
the radius of the Keplerian orbit is estimated as $\sim$100 $r_g$ , equivalently, $\sim$0.005 parsecs. Now this result can be interpreted in terms of binary supermassive black hole system \citep[e.g.,][]{Valtonen2011}, which potentially can explain presence of year-like periodic timescales in AGN \cite[see for a review][]{Komossa2006}.  For the binary mass ratios in the rage 0.1--0.01 \cite[see e.g.,][]{Begelman1980,Sillanpaa1988}, the orbital decay timescale in the GW-driven regime can be computed as   
 \begin{equation}
\tau_{insp}=3.05\times 10^{-6} \left ( \frac{M}{10^9M_\odot} \right )^{-3}\left ( \frac{a}{r_g} \right )^4 \rm years,
\end{equation}
\noindent \cite[see][]{Peters1964}; the timescale turns out to be less than a thousand years.  Indeed, existence of binary SMBH is consistent with the prediction by the hierarchical galaxy formation models. As a matter of fact, the closest binary SMBH so far detected lies within a sub-parsec separation ($\sim$0.35 pc; \cite{Kharb2017}) in Mrk 533, a  Seyfert type 2 AGN. Therefore if the above periodicity results from such a system, we should expect that the system will undergo gravitational coalescence within a few centuries accompanied with the emission of gravitational waves of the frequency $\sim10^{-2}$ $ \mu$Hz. However, the probability of observing such a close (milliparsec) system of binary SMBH might be too small \cite[see ][]{Ackermann2015}. But on the other hand,  in the case of strong disk-jet connection \cite[e.g., in][]{Blandford1982} (see also \cite[][]{Bhatta2018c}), the periodic modulations induced due to  binary SMBH system could propagate down the jet affecting the Doppler boosted emission. In such case, true periodic timescales at the BH could be longer than the observed one due to relativistic effects, i.e.  $P=\delta /(1+z)P_{obs}$, where Doppler factor, $\delta =(\Gamma \left ( 1-\beta cos\theta \right ))^{-1}$,  and  $\Gamma$, $\beta=v/c$ and $\theta$ are the bulk Lorentz factor, relativistic speed and  angle  to the line of sight, respectively. Consequently, the corresponding Keplerian orbit of the secondary black hole could be larger and therefore lie farther away from the primary BH. For instance, if a jet component carrying the periodic signal travels along the jet viewed at $\sim1^o$ with a bulk Lorentz factor of $\sim15$, the orbit estimated using dilated timescale gives a distance of a $\sim0.1$ parsecs. Such a sub-parsec separation incidentally represents a relatively stable configuration in the evolution of binary SMBH systems \cite[see][]{Rieger2007}. The left panel of Figure \ref{Fig4} shows the radii of the  Keplerian circular orbits of the possible secondary black hole corresponding to periodic modulations propagating with three different bulk Lorentz factors, $\Gamma =10, 15$, and 20, along the blazar jet that is viewed within $ 10^o$.
 
 Similarly, the periodic changes can also be associated with relativistic motion of the emission regions along the helical path of the magnetized jets \cite[e.g.,][]{Camenzind92, Mohan2015}.  In particular,  when the emission regions move along the helical path of a jet with a  high bulk Lorentz factor, due to the relativistic effects, the periodic changes in the viewing angle cause the  Doppler boosted emission to be periodically modulated. The rest frame flux (${F}'_{{\nu'}}$) and observed flux ($F_{\nu}$) are related through the equations
  \begin{equation}
F_{\nu}(\nu)=\delta(t)^{3+\alpha}{F}'_{{\nu}'} (\nu)   \quad\text{and}\quad      \delta(t)=1/\Gamma \left ( 1-\beta cos\theta(t)  \right )
\label{flux}
\end{equation}
 
\noindent \citep[similar to Equation B4 in][]{Urry1995}.  For blazars having typical radio spectral index ($\alpha=0.6$),  the change in the angle of line of sight required to double the observed flux, while keeping the flux in the source rest frame constant, for various jet angles and three bulk Lorentz factors $\Gamma=10,\ 15\ \rm and\ 20$, is displayed  in the right panel of Figure \ref{Fig4}. As the figure shows, for typical blazar viewing angles in the $1-5 ^o$ range, a slight change in the viewing angle e.g., $\sim 1.5^o$, is sufficient to double the observed brightness.

 In another likely scenario, the detected QPO  with a periodicity of $\sim 563$\,days could be explained in the context of  instabilities intrinsic to blazar accretion disk.  For example, bright hot-spots on the disk that revolve around the BH could result in the year timescale periodicity. Assuming they follow circular Keplerian orbits, they could be located at  $\sim100\ r_g$ from the BH. In the General Relativistic treatment of rotating massive objects, the nearby inertial frames are distorted by frame-dragging resulting in the nodal precession of the tilted orbits, {\b known as the Lense-Thirring precession}. Periodic behaviors in astronomical systems can also be linked to the Lense-Thirring precession of accretion disks \citep[e.g.][]{Stella1998,Motta2011}. The timescale of such precession is proportional to the distance (from the BH) cubed and also to the mass of the central black hole so that the timescale, $\tau_{LT}$, can be expressed as 
 \begin{equation}
\tau_{LT}=0.18 \left ( \frac{1}{a_s} \right )\left ( \frac{M}{10^9M_\odot} \right )\left ( \frac{r}{r_g} \right )^{3} \rm days,
\end{equation}
\noindent where $a_s$, $M$ and $r$ represent dimensionless spin parameter, mass of the BH and the radial distance of the emission region from the BH, respectively. For a maximally spinning ($a_s=0.9$) central black hole  with a mass ${10^9M_\odot}$, a timescale of 360 days places the emission region around $\sim 12\ r_g$. The periodic oscillations could  be the result of the jet precession due to such warped accretion disks.  However, such a periodic timescale would be thousands of years, hence may not be relevant here \cite[][]{Liska2018,Graham2015}. 
 In radio-loud  AGNs, magnetic flux accumulation in the accretion disk can result in the formation of so-called \emph{magnetically choked accretion flow}. In that case, owing to sudden change in the density and the magnetic flux, disk instabilities e.g., the Rayleigh-Taylor and Kelvin-Helmholtz instabilities, are set up; this in turn can induce QPOs  at the  disk-magnetosphere interface \citep{Li2004,Fu2012}. Similar QPOs have been observed in the recent magneto-hydrodynamical  simulations of the large scale jets \cite{McKinney2012}.  The periodic timescale of  these QPOs could range from a few days to a few years depending upon the black hole mass and the spin parameter. 
 
 \section{Conclusion}
 
The long term ($\sim$10.5 years) radio (15 GHz) observations of the blazar  J1043+2408 were analyzed for possible periodicities using three methods widely used in  astronomical time series  analysis: epoch folding, Lomb-Scargle periodogram and auto-correlation function. The study revealed a strong periodic signal with a $\sim$ 560 day periodicity. A large number of Monte Carlo simulation of the light curves were used to establish a high significance ($ > 99\%$) of the signal against possible spurious detection. We conclude that while other above discussed scenarios can not be completely ruled out, periodic modulations induced by gravitational perturbation in binary SMBH system seems a more plausible mechanism at the root of the observed periodic radio signal.

%%%%%%%%%%%%%%%%%%%%%%%%%%%%%%%%%%%%%%%%%%
\vspace{6pt} 

%%%%%%%%%%%%%%%%%%%%%%%%%%%%%%%%%%%%%%%%%%
\acknowledgments{We acknowledge the financial support by the Polish National Science Centre through the grant UMO-2017/26/D/ST9/01178. This research has made use of data from the OVRO 40-m monitoring program (Richards, J. L. et al. 2011, ApJS, 194, 29) which is supported in part by NASA grants NNX08AW31G, NNX11A043G, and NNX14AQ89G and NSF grants AST-0808050 and AST-1109911. We thank Prof. James Webb for carefully reading the manuscript. The  author would like to thank the anonymous referees for their constructive comments and suggestions that greatly improved the manuscript.}

%%%%%%%%%%%%%%%%%%%%%%%%%%%%%%%%%%%%%%%%%%
\conflictsofinterest{The authors declare no conflict of interest} 

%%%%%%%%%%%%%%%%%%%%%%%%%%%%%%%%%%%%%%%%%%
%% optional
\abbreviations{The following abbreviations are used in this manuscript:\\

\noindent 
\begin{tabular}{@{}ll}
AGN & Active Galactic Nuclei\\
ACF & Auto-correlation function\\
BL Lac & BL Lacertae object\\
BH & Black hole\\
FSRQ & Flat Spectrum Radio Quasar\\
LSP & Lomb-Scargle Periodogram\\
MC & Monte Carlo\\
OVRO&Owens Valley Radio Observatory \\
PSD&Power spectral density \\
QPO&Quasi-periodic oscillation \\
SMBH& Supermassive black hole\\
\end{tabular}}

% References, variant A: internal bibliography
%=====================================
\reftitle{References}

% The following MDPI journals use author-date citation: Arts, Econometrics, Economies, Genealogy, Humanities, IJFS, JRFM, Laws, Religions, Risks, Social Sciences. For those journals, please follow the formatting guidelines on http://www.mdpi.com/authors/references
% To cite two works by the same author: \citeauthor{ref-journal-1a} (\citeyear{ref-journal-1a}, \citeyear{ref-journal-1b}). This produces: Whittaker (1967, 1975)
% To cite two works by the same author with specific pages: \citeauthor{ref-journal-3a} (\citeyear{ref-journal-3a}, p. 328; \citeyear{ref-journal-3b}, p.475). This produces: Wong (1999, p. 328; 2000, p. 475)

%=====================================
% References, variant B: external bibliography
%=====================================
%\externalbibliography{yes}
%\bibliography{your_external_BibTeX_file}

%%%%%%%%%%%%%%%%%%%%%%%%%%%%%%%%%%%%%%%%%%
%% optional
\sampleavailability{Samples of the compounds ...... are available from the authors.}

%% for journal Sci
%\reviewreports{\\
%Reviewer 1 comments and authors’ response\\
%Reviewer 2 comments and authors’ response\\
%Reviewer 3 comments and authors’ response
%}

%%%%%%%%%%%%%%%%%%%%%%%%%%%%%%%%%%%%%%%%%%
\end{document}